\documentclass[
    ,final            
  ]
  {aipproc}

\layoutstyle{6x9}

\begin{document}

\title{Singularity Avoidance in Nonlinear Quantum Cosmology}


\author{Le-Huy Nguyen}
  {address={Department of Physics, National University of Singapore, Kent Ridge, Singapore.}}

\author{Rajesh R. Parwani\footnote{parwani@nus.edu.sg}}
{address={Department of Physics, National University of Singapore, Kent Ridge, Singapore.}}

\begin{abstract}
 We extend our previous study on the effects of an information-theoretically motivated nonlinear correction to the Wheeler-deWitt equation in the minisuperspace scheme for FRW universes. Firstly we show that even when the geometry is hyperbolic, and matter given by a cosmological constant, the nonlinearity can still provide a barrier to screen the initial  singularity, just as in the case for flat universes. Secondly, in the flat case we show that  singularity avoidance in the presence of a free massless scalar field is perturbatively possible for a very large class of initially unperturbed quantum states, generalising our previous discussion using Gaussian states. 
\end{abstract}

\maketitle


\section{Introduction}
One motivation for studying quantum cosmology \cite{Kiefer} is to examine if the universe  can have a nonsingular evolution when extrapolated back in time. The standard Wheeler-deWitt (WDW) approach does not always succeed in this respect while, for example, a modified loop quantisation seems promising \cite{Ash}.
However as the early quantum universe is still very much an area of theoretical speculation, it is useful to attempt other approaches which might provide different signatures for future experimental tests of possible late time effects of the early time modifications.

In Ref.\cite{LP} we studied a nonlinear modification of the WDW equation in the the minisuperspace scheme. Our nonlinearisation was exactly the one introduced  in \cite{RP} on the basis of the maximum entropy (uncertainty) method. We refer the interested reader to Refs.\cite{RP,LP} for the detailed motivation.

As in \cite{LP} we assume here that the nonlinearity is weak so that perturbative methods are applicable.

  \section{$k=-1$, Cosmological Constant case}
Consider the Einstein-Hilbert action for a  FRW universe,
\begin{equation}
S=\int dt L = {1 \over 2} \int dt N a^3 \left[ {-\dot{a}^2 \over N^2 a^2} + 
{\dot{\phi}^2 \over N^2} -V(\phi) + {k \over a^2} \right]
\end{equation}
where $a(t)$ is the dimensionless scale factor, $V(\phi)$ the potential energy of the scalar field $\phi(t)$, $N$ the lapse function, $k=0,\pm 1$, and we have taken $\hbar=c=1$.

 If matter is given by a cosmological constant, set $1/V = a_{0}^2$, so that the action is
\begin{equation}
S=\int dt L = {1 \over 2} \int dt N \left[ {-\dot{a}^2 a\over N^2 } + a(k- {a^2 \over a_{0}^2} ) \right] \label{frw1} \, ,
\end{equation}
 the cosmological constant $\Lambda =3/a_{0}^2$ modelling inflationary sources in the early universe. The Friedmann equation is
  \begin{equation}
 \dot{a}^2 + (k-a^2/a_{0}^2) =0 \, , \label{fried}
\end{equation}
 while standard quantisation \cite{Atkatz} gives the corresponding WDW equation in minisuperspace 
\begin{equation}
\left[ -{\partial^2 \over \partial a^2} + V(a) \right] \psi(a) =0 \, , \label{wdw1}
\end{equation}
which is a one-dimesional, $0 \le a < \infty$, time-independent Schrodinger equation 
for a particle moving in a potential $V(a)=a^2(k-a^2/a_{0}^2)$. 

As in Ref.\cite{LP} we model the unknown new physics as short distances by a nonlinear correction, first introduced in Ref.\cite{RP}, so that the modified equation becomes 
\begin{equation}
\left[ -{\partial^2 \over \partial a^2} + V(a) + F(p) \right] \psi(a) =0 \label{nl1}
\end{equation}
where
\begin{eqnarray} 
  F(p) &\equiv& Q_{1NL} - Q \, , \label{F} 
\end{eqnarray}
with 
\begin{equation}
Q_{1NL}= {  1  \over 2 L^2 \eta^4}  \left[ \ln {p \over (1-\eta) p + \eta p_{+} } + 1 - {(1-\eta) p \over (1-\eta) p + \eta p_{+}} - {\eta p_{-} \over (1-\eta) p_{-} + \eta p} \right]  \,  \label{Q2}
\end{equation}
and 
\begin{eqnarray}
Q &=&  -  {1 \over \sqrt{p}} {\partial^2 \sqrt{p} \over \partial a^2} \, \; . \label{pot1} 
\end{eqnarray}
Here $p(a) = \psi^{\star}(a) \psi(a)$ and $p_{\pm}(a)  \equiv  p(a \pm \eta L)$.
 The variable $a$ is dimensionless, $L>0$ is the dimensionless nonlinearity scale and $0 < \eta < 1$ is a parameter that labels a family of nonlinearisations.

Setting $k=-1$  in (\ref{nl1}) we get 
\begin{equation}
\left[ -{\partial^2 \over \partial a^2} -a^2 -{a^4 \over a_{0}^2} +  F(p)) \right] \phi(a) =0 \, .
\end{equation}
We assume, as in our previuous $k=0$ study, that the nonlinearity is small even in the early universe so that $F$ may be expanded perturbatively in $L$ to lowest order. We also assume now that in the region of interest, the scale factor is of order $\sqrt{L}$ so that the $a^4$ term may be ignored compared to the $a^2$ term. Thus with
\begin{equation}
f(a)={p' \over 12 p^3} (2 p'^2 -3p''p) \,  \label{f1est}
\end{equation}
we may write 
\begin{equation}
\left[ -{\partial^2 \over \partial a^2} -a^2 + \eta(3-4\eta) L f(a) \right] \phi(a) =0 \, . 
\end{equation}
We solve the equation by iterating about the unperturbed solution for $L=0$. The unperturbed solution which represents an expanding universe at large scale is given by the Hankel function
\begin{eqnarray}
\phi_0(a) &\propto&  \sqrt{a} H_{1/4}^{(2)} (a^2/2) \, .
\end{eqnarray}
From this unpertubed solution one calculates $p_0=\phi_{0}^{*} \phi$ and thus to leading nontrivial order in the iteration one has a linear Schrodinger equation with an effective potential 
\begin{equation}
V_{eff} = -a^2 + \eta(3-4\eta) L f_0(a)  \, .
\end{equation}
For small $a$, 
\begin{equation}
f_0 \approx 0.07 +0.2a \label{f0}
\end{equation}
and so, as in the case for $k=0$, for $\eta <3/4$ there is an effective potential barrier that screens the original classical singularity, a finite size universe coming into being through quantum tunneling. 

The probability for tunneling through the barrier is given in the WKB approximation by 
an expression similar to the $k=0$ case,
\begin{equation}
P_{k=-1} \approx \exp{\left[-0.1 \eta (3-4\eta) L) \right]} \, ,
\end{equation}
so that for fixed $\eta < 3/4$  small values of $L$ are ``preferred", self-consistent with our approximation  $L \ll 1$.

This exercise shows that even in the regime where the nonlinearity is purely perturbative, a barrier screens the initial singularity in the hyperbolic geometry for {\it the same range of $\eta <3/4$} as for the case of flat geometry discussed in \cite{LP}.

\section{$k=0$, Free Massless Scalar Field}
In terms of $\alpha = \ln a$ the $k=0$ classical FRW action becomes
\begin{equation}
S=\int dt L = {1 \over 2} \int dt \ N e^{3 \alpha} \left[ {-\dot{\alpha}^2 \over N^2 } + 
{\dot{\phi}^2 \over N^2}  \right] \, ,  \label{act2}
\end{equation}
giving the following classical equations in the $N=1$ gauge,
\begin{eqnarray}
\ddot{\phi} + 3 \dot{\phi} \dot{\alpha} &=& 0 \, , \\
2 \ddot{\alpha} + 3 \dot{\alpha}^2 + 3\dot{\phi}^2 &=& 0 \, , \\
-\dot{\alpha}^2 + \dot{\phi}^2 &=& 0 \, . \label{const1}
\end{eqnarray}

The coresponding WDW equation equation is now the Klien-Gordon equation
\begin{equation}
\left[-{\partial^2 \over \partial \phi^2}+{\partial^2 \over \partial \alpha^2} \right] \psi(\phi, \alpha) =0 \, . \label{wdw2}
\end{equation}
The general solution of (\ref{wdw2}) is of course given by $\psi(\phi,\alpha) = f_1(\phi - \alpha) + f_2(\phi + \alpha)$ where the $f_i$'s are any differentiable functions.. We will choose a state which is localised at large values of $\alpha$ at large intrinsic time $\phi$, corresponding to a large universe at large times. Thus we take the density $p(\phi, \alpha) \equiv \psi^{*} \psi \equiv h(z^2)$ where $z=\phi-\alpha$. {\it The function $h$ is taken to be positive, localised, with a maximum at $z=0$, normalizable, but otherwise arbitrary}. Note that $-\infty<z<\infty$.

We will use that initial state, given by $h$, to solve by perturbation and iteration the nonlinearised Klien-Gordon equation which follows from the information-theoretic approach discussed in Ref.\cite{LP}. To lowest nontrivial order in the nonlinearity scale $L$ we get an effective linear equation (see \cite{LP} for details in the case of initial Gaussian states)
\begin{equation}
\left[-{\partial^2 \over \partial \phi^2}+{\partial^2 \over \partial \alpha^2}  + V_{eff} \right]  \psi(\phi, \alpha) =0 \,  \label{eff2}
\end{equation}
with 
\begin{eqnarray}
V_{eff} &=& u \left[ {2 z h' \over h} \right] \left[ 2 \left( {2 z h' \over h} \right)^2 -3 \left( {2 h' \over h} + {4 z^2 h'' \over h }\right) \right] \, , \label{veff} \\
u &\equiv& { \eta (3-4\eta) (L_{\alpha} + L_{\phi} )  \over 12} \, ,
\end{eqnarray}
where the generally distinct parameters $L_{\alpha}>0$ and $L_{\phi}>0$ correspond to the nonlinear effects in the gravitational and matter degrees of freedom. The prime, $'$, symbol means ${\partial \over \partial z^2}$.

The effective classical equations that imply, through the correspondence principle, the modified quantum equation (\ref{eff2}) are given in the $N=1$ gauge by
\begin{eqnarray}
\ddot{\phi} + 3 \dot{\phi} \dot{\alpha} + {1 \over 2 } e^{-6\alpha} {\partial V_{eff} \over \partial \phi} &=& 0 \, , \label{m1} \\
2 \ddot{\alpha} + 3 \dot{\alpha}^2 + 3\dot{\phi}^2 + e^{-6\alpha} \left[ 3 V_{eff} - {\partial V_{eff} \over \partial \alpha} \right] &=& 0 \, , \label{m2} \\
-\dot{\alpha}^2 + \dot{\phi}^2 +  e^{-6\alpha} V_{eff} &=& 0 \, . \label{const2}
\end{eqnarray}
Note that 
\begin{equation}
{\partial V_{eff} \over \partial \phi} = -{\partial V_{eff} \over \partial \alpha}.
\end{equation}

\subsection{Analytical Results}
The coupled equations (\ref{m1}-\ref{const2}) have solutions for which $\alpha =\phi$ at all times; the constraint equation (\ref{const2}) is then automatically satisfied and the other equations reduce to 
\begin{equation}
3 \dot{\alpha}^2 + \ddot{\alpha} -{1 \over 2} e^{-6 \alpha} \left({\partial V_{eff} \over \partial \alpha}\right)_{\alpha=\phi}  =0 \, . \label{exact}
\end{equation}
Now writing $h(z^2) = a + bz^2 + cz^4$ in the neighbourhood of $z=0$, and using (\ref{veff}) we get
\begin{equation}
B \equiv \left({\partial V_{eff} \over \partial \alpha}\right)_{\alpha=\phi} = 3u\left({2b \over a}\right)^2
\end{equation}
which is positive for $\eta < 3/4$. Eq.(\ref{exact}) can be integrated to give 
\begin{equation}
e^{6 \alpha} \ \dot{\alpha}^2 = B \alpha + C \label{analy}
\end{equation}
where $C$ is the integration constant. Thus for $\eta < 3/4$
\begin{equation}
\alpha \ge {- C \over B}  \label{amin}
\end{equation}
and so 
\begin{equation}
a \ge a_{min} = \exp({-C \over B}) \, .
\end{equation}

That is, for the subset of solutions for which $\alpha(t) \equiv \phi(t)$ for all $t$, the universes have a minimum nonzero size, which depends on the initial conditions that fix $C$: The Big Bang singularity is avoided, being replaced by a bounce. 

These $\alpha \equiv \phi$ solutions of the modified dynamics are stable to small perturbations as one can see by setting $\alpha(t) = \phi + \delta(t)$ and linearising the dynamical equations for $\delta \ll 1$ to get
\begin{equation}
\ddot{\delta} + {9 u \delta \gamma \over 2 a^6(t) } =0 \, , \label{sho}
\end{equation}
where $\gamma = (2 h'/h)^2$ evaluated at $z=0$. Since (\ref{sho}) is the equation for an oscillator, $\delta$ is bounded.

The main point to note is that while a $\alpha \equiv \phi$ correlation is also possible  in the original classical dynamics, only in the modified dynamics induced by the quantum nonlinearity is the singularity avoided. 

We have not analysed the modified classical dynamics for $\alpha \neq \phi$ cases when the  initial state $h$ is not a Gaussian. For Gaussian $h$ the numerical analysis for $\alpha \neq \phi$ cases was reported in \cite{LP}, showing a bounce in the regime where the nonlinearity was self-consistently small (in accordance with our perturbative approximation which led to the modified dynamics). We expect similar results for non-Gaussian $h$ discussed in this paper though this needs verification.

\section{Outlook}
The main result here has been to generalise the  analysis for the $k=0$ geometry with free massless scalar field matter discussed in \cite{LP}. We have shown that (for the $\alpha =\phi$ subset and perturbations thereof) the singularity is again perturbatively avoided, being replaced by a bounce, for initial localised states which are more general than a Gaussian
wavepacket.

As metioned in \cite{LP}, the next step would be to study the nonlinearity non-perturbatively to see if the initial singularity is also avoided when the nonlinearity is large and if realistic inflation can be generated by the nonlinearity starting from free matter fields. Other initial situations worthy of study are:  more general matter and less symmetric geometries.


\begin{theacknowledgments}
  RP thanks the organisers for a stimulating conference and their wonderful hospitality.

\end{theacknowledgments}

\end{document}